\documentclass[letterpaper, 10 pt, conference]{ieeeconf}

\usepackage{graphicx}
\usepackage{epstopdf}
\usepackage{epsfig}
\usepackage{times}
\usepackage{amsmath}
\usepackage{amssymb}
\usepackage{latexsym}
\usepackage{color}
\usepackage{verbatim}
\usepackage{cite}
\usepackage{BernsteinStyle}
\usepackage[font = footnotesize]{caption}
\usepackage[font = footnotesize]{subcaption}
\usepackage[hidelinks]{hyperref}
\usepackage{makecell}

\makeatletter
\newcommand*\bigcdot{\mathpalette\bigcdot@{.5}}
\newcommand*\bigcdot@[2]{\mathbin{\vcenter{\hbox{\scalebox{#2}{$\m@th#1\bullet$}}}}}
\makeatother

\IEEEoverridecommandlockouts

\title{\LARGE
Target Tracking and Prediction in the Frenet-Serret Frame\\ Using Curvature and Torsion Estimation
}

\title{\LARGE
Target Tracking and Prediction \\ Based on Curvature and Torsion Estimation
}

\title{\LARGE
Target-Trajectory Prediction Using Adaptive Numerical Differentiation\\
for Curvature and Torsion Estimation
}

\title{\LARGE
Target-Trajectory Prediction Using Curvature and Torsion Estimation
}

\title{\LARGE
Target Trajectory Prediction Using Curvature and Torsion Estimation
}

\title{\LARGE
Frenet-Serret-Based Trajectory Prediction
}

\author{Shashank Verma and Dennis S. Bernstein%
\thanks{$^{*}$Shashank Verma and Dennis S. Bernstein are with the Department of Aerospace Engineering, University of Michigan, Ann Arbor, MI 48109, USA 
{\tt\small shaaero@umich.edu}}%
}

\begin{document}

\maketitle
\thispagestyle{empty}
\pagestyle{empty}

\begin{abstract}
Trajectory prediction is a crucial element of guidance, navigation, and control systems. 
This paper presents two novel trajectory-prediction methods based on real-time position measurements and adaptive input and state estimation (AISE). 
The first method, called AISE/va, uses position measurements to estimate the target velocity and acceleration. 
The second method, called AISE/FS, models the target trajectory as a 3D curve using the Frenet-Serret formulas, which require estimates of velocity, acceleration, and jerk. 
To estimate velocity, acceleration, and jerk in real time, AISE computes first, second, and third derivatives of the position measurements.
AISE does not rely on assumptions about the target maneuver, measurement noise, or disturbances. 
For trajectory prediction, both methods use measurements of the target position and estimates of its derivatives to extrapolate from the {current} position. 
The performance of AISE/va and AISE/FS is compared numerically with the $\alpha$-$\beta$-$\gamma$ filter, which shows that AISE/FS provides more accurate trajectory prediction than AISE/va and traditional methods, especially for complex target maneuvers. 
\end{abstract}

\section{INTRODUCTION}
Trajectory prediction is a crucial element of guidance, navigation, and control systems. 
This objective requires estimation algorithms that account for measurement noise and changing environmental conditions
%
\cite{zarchan_book_2012,BarShalom2001Estimation}.
Trajectory-prediction methods for aircraft can be categorized into three main areas: state estimation models, kinematic models, and machine learning-based models \cite{zeng_aerospace_aircraft_4D_2022_review}. 
For autonomous vehicles, a survey of trajectory-prediction methods 
is given in  \cite{Huang_2022_traj_pred_automobile}.

State-estimation-based trajectory prediction has been widely studied. 
Chatterji \cite{chatterhi_short_term_1999} used a Kalman filter (KF) for short-term prediction by estimating ground speed and trajectory angle, and then propagating the position using kinematic equations. 
However, the reliance of this method on fixed aircraft intention affects accuracy when deviations occur. 
To address inefficiencies in high-dimensional state estimation, Lymperopoulos et al \cite{Lymperopoulos_MC_trajPredic_2010} proposed a particle filter as an alternative to sequential Monte Carlo techniques, improving trajectory prediction under complex conditions. 

Trajectory prediction is vital for air traffic management (ATM) and flight planning. Ayhan and Samet \cite{ayhan_predictive_analy_2016} introduced a stochastic model using 3D grids and weather data, improving safety, efficiency, and fuel savings in ATM.
Similarly, Lin et al \cite{lin_HMM_2018} proposed a method based on relative motion between positions using historical data, hidden Markov models, and Gaussian mixture models to enhance prediction accuracy.
%
%
%

%
Ammoun \cite{ammoun_rt_prediction_2009} uses the Kalman filter for real-time trajectory prediction. Lefkopoulos \cite{Lefkopoulos_MMFK_2021} proposed interaction-aware motion prediction for autonomous driving, which is based on the interacting-multiple-model Kalman filter. 
%
%
 

%
Input-estimation techniques for maneuvering targets were explored by Lee \cite{hungu_lee_generalized_1999} and Bar-Shalom \cite{shalom-1989-input-est}, and a Kalman-filter-based scheme incorporating input estimation was introduced by Khaloozadeh \cite{khaloozadeh_modified_2009}. 
Gupta \cite{gupta_retrospective-cost-based_2012}, Ahmed \cite{ahmed_input_2019}, and Han \cite{han_rcie}  extended these methods by introducing adaptive input-estimation techniques for maneuvering targets.
Additionally, Tenne \cite{tenne-2002-alpha-beta-gamma} analyzed the performance of $\alpha$-$\beta$-$\gamma$ filters for constant-acceleration targets, while Hasan \cite{hasan-2013-adaptive-alpha-beta} proposed an adaptive $\alpha$-$\beta$ filter using genetic algorithms for real-time parameter adaptation.

Machine learning has increasingly been integrated into predictive models for trajectory prediction, incorporating neural networks and data-driven approaches. 
Akcal et al \cite{akcal_predictive_2021} introduced a recurrent neural network to predict target acceleration within a pursuer guidance algorithm.
%
%
Pang et al \cite{yutian_2020_probalistic} applied a Bayesian neural network for probabilistic trajectory prediction, using approximate Bayesian inference, which generated trajectory predictions with confidence intervals.

More recent advancements leverage differential geometry for target tracking, particularly in 3D maneuvering environments. 
Bonnabel et al \cite{bonnabel_cdc_2017_target_tracking} introduced a method using the Frenet-Serret frame to track target motion based on position measurements, under the assumption of constant speed, uniform curvature, and planar motion. 
%
%
%
%
%
Further extensions of the Frenet-Serret framework by Gibbs \cite{gibbs_fs_iekf_2022} used IEKF to track accelerating targets. 
Giulio et al \cite{giulio_2004_frenet} used a Frenet-Serret-based trajectory-prediction method with accelerometer and rate-gyro data, estimating curvature and torsion parameters of the Frenet frame for dynamic environments.
These methods differ from the current work, where the focus is on predicting target trajectories using only position measurements.
%
%
%

%
Two novel methods for trajectory prediction are presented. The first method estimates the target velocity (v) and acceleration (a) by estimating the first and second derivatives of the position measurements; this approach is called  AISE/va. 
The second method models the target trajectory as a three-dimensional curve using the Frenet-Serret (FS) formulas, which require estimates of the velocity, acceleration, and jerk of the target position; this method is called AISE/FS.
For real-time derivative estimation, adaptive input and state estimation (AISE) is used \cite{verma_shashank_2023_realtime_IJC,verma_shashank_2024_realtime_VRF_axiv,verma_shashank_ACC2022}. 
AISE operates without assumptions about the target maneuver, measurement noise, or disturbances, thereby eliminating the need for prior information about the target or sensor characteristics. 
{For trajectory prediction, both methods use measurements of the target position and estimates of its derivatives to extrapolate from the current position.} 
A summary of AISE is given in Section \ref{sec:AISE}.
The performance of AISE/va and AISE/FS is compared numerically with the $\alpha$-$\beta$-$\gamma$ filter, which shows that AISE/FS provides more accurate trajectory prediction than AISE/va and traditional methods, especially for complex target maneuvers. 

This paper is organized as follows: Section \ref{sec:prob_statement} introduces the problem statement. Section \ref{sec:AISE} discusses the AISE method. Section \ref{sec:aise_va} introduces AISE/va, and Section \ref{sec:aise_fs}  presents AISE/FS for trajectory prediction. Finally, Section \ref{sec:num_example} presents two numerical examples comparing these methods with conventional approaches.


\section{Problem Statement}  \label{sec:prob_statement}

We assume that the Earth is inertially non-rotating, and non-accelerating.
The right-handed frame $\rmF_{\rm E} = \begin{bmatrix} \hat{\imath}_{\rmE} & \hat{\jmath}_{\rmE} & \hat{k}_{\rmE} \end{bmatrix}$ is fixed relative to the Earth, with the origin $\rmo_{\rmE}$ located at any convenient point on the Earth's surface;
hence, $\rmo_{\rmE}$ has zero inertial acceleration.
$ \hat{k}_{\rmE}$ points downward, and $\hat{\imath}_{\rmE}$ and $\hat{\jmath}_{\rmE}$ are horizontal. 
%
$\rmo_{\rm T}$ is any point fixed on the target.

\begin{figure}[h!t]
              \begin{center}
            {\includegraphics[width=0.8\linewidth]{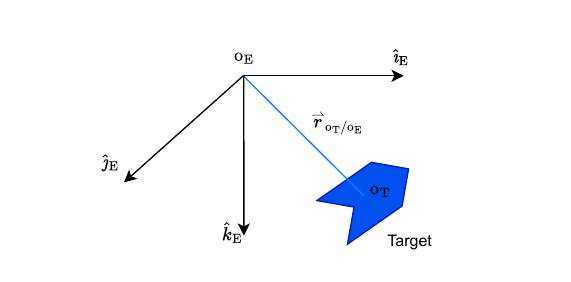}}
            \end{center}
            \caption{ $\vect{r}_{\rmo_{\rm T}/\rmo_{\rmE}}$ is the physical position vector between $\rmo_{\rm T}$ and $\rmo_\rmE$. } 
            \label{target_fig}
\end{figure}

The location of the target origin $\rmo_{\rm T}$ relative to $\rmo_{\rmE}$ at each time instant is represented by the position vector $\vect{r}_{\rmo_{\rm T}/\rmo_{\rmE}}$, as shown in Figure \ref{target_fig}. 
We assume that a sensor measures the position of the target in the frame $\rmF_{\rm E}$ as
\begin{align}
    p_k \isdef \vect{r}_{\rmo_{\rm T}/\rmo_{\rmE}} (t_k) \bigg\vert_\rmE \in \mathbb{R}^3, \quad  
     \label{pos_resolve_E}
\end{align}
where $k$ is the step and $t_k \isdef kT_\rms$. Here, $p_{k} = p(kT_{\rms})$ is the position measurements of the target at step $k$, with $T_{\rms}$ being the sample time.
The first and second derivatives of $\vect{r}_{\rmo_{\rm T}/\rmo_{\rmE}}$ with respect to $\rmF_{\rm E}$ represent the physical velocity and acceleration vector $\framedotE{\vect{r}}_{\rmo_{\rm T}/\rmo_{\rmE}}$ and $\frameddotE{\vect{r}}_{\rmo_{\rm T}/\rmo_{\rmE}}$. Resolving $\framedotE{\vect{r}}_{\rmo_{\rm T}/\rmo_{\rmE}}$ and $\frameddotE{\vect{r}}_{\rmo_{\rm T}/\rmo_{\rmE}}$ in $\rmF_{\rm E}$ yields
\begin{align}
v_k \isdef \dot{p}_k = 
 \,\,\framedotE{\vect{r}}_{\rmo_{\rm T}/\rmo_{\rmE}}(t_k)\bigg\vert_\rmE \in \mathbb{R}^3, 
\label{vel_resolve_E}
\end{align}
\begin{align}
a_k \isdef \ddot{p}_k =  \, \frameddotE{\vect{r}}_{\rmo_{\rm T}/\rmo_{\rmE}}(t_k)\bigg\vert_\rmE \in \mathbb{R}^3 .  \label{acc_resolve_E}
\end{align}



Using position measurements of the target given by \eqref{pos_resolve_E} at step $k$,  we predict the future trajectory $\hat{p}_{k+l}$ for all $l=1,\ldots,\ell$.

\section{Review of Adaptive Input and State Estimation} \label{sec:AISE}
Here we summarize AISE \cite{verma_shashank_2023_realtime_IJC,verma_shashank_2024_realtime_VRF_axiv,verma_shashank_ACC2022} for real-time numerical differentiation.  
%
%
Consider the linear discrete-time SISO system
\begin{align}
	x_{k+1} &=  A x_{k} + Bd_{k}, 	\label{state_eqn}\\
	y_k  &= C x_k + D_{2,k} w_k, \label{output_eqn}
\end{align}
where
$k\ge0$ is the step,
$x_k \in \mathbb R^{n}$ is the unknown state,
$d_k \in \mathbb R$ is unknown input,
$y_k \in \mathbb R$ is a measured output,
$w_k \in \mathbb R$ is standard white noise, 
and $D_{2,k}w_k \in \mathbb R$ is the measurement noise at time $t = kT_\rms$, where $T_\rms$ is the sample time.
The matrices $A \in \mathbb R^{n \times n}$, $B \in \mathbb R^{n \times 1}$, and $C \in \mathbb R^{1 \times n}$ are assumed to be known, and $D_{2,k}$ is assumed to be unknown.
The sensor-noise covariance is $V_{2,k} \isdef D_{2,k} D_{2,k}^\rmT$.
The goal of adaptive input estimation (AIE) is to estimate $d_k$ and $x_k$.

To apply AIE to real-time numerical differentiation, we use \eqref{state_eqn} and \eqref{output_eqn} to model a discrete-time integrator. As a result, AIE provides an estimate $\hat{d}_k$ of the derivative of the sampled output $y_k$. 
For single discrete-time differentiation,  
\begin{align}
A = 1, \quad B = T_\rms, \quad C=1, \label{single_integrator}
\end{align}
for double discrete-time differentiation,  
\begin{align}
    A = \begin{bmatrix}
        1 & T_\rms\\ 0 & 1
    \end{bmatrix}, \quad B = \begin{bmatrix}
        \half T_\rms^{2} \\ {T_\rms}
    \end{bmatrix}, \quad C = \begin{bmatrix}
        1 & 0
    \end{bmatrix}, \label{double_integrator}
\end{align}
and  for triple discrete-time differentiation, 
\begin{align}
    A = \begin{bmatrix}
        1 & T_\rms & \half T_\rms^{2}\\ 0 & 1 & T_\rms \\ 0 & 0 & 1
    \end{bmatrix}, \quad B = \begin{bmatrix}
        \frac{1}{6}T^3_\rms \\[.5ex]  \half T_\rms^{2} \\ {T_\rms}
    \end{bmatrix}, \quad C = \begin{bmatrix}
        1 & 0 & 0
    \end{bmatrix}. \label{triple_integrator}
\end{align}
{Note that \eqref{double_integrator} represents a discretized double integrator. Therefore, the output of \eqref{double_integrator} is approximately the second integral of the input of \eqref{double_integrator}. Equivalently, the input of \eqref{double_integrator} is approximately the second derivative of the output of \eqref{double_integrator}. Similar statements hold for \eqref{single_integrator} and \eqref{triple_integrator}.
}


\subsection{Input Estimation}
AIE comprises three subsystems, namely, the Kalman filter forecast subsystem, the input-estimation subsystem, and the Kalman filter data-assimilation subsystem.
First, consider the Kalman filter forecast step
%
\begin{gather}
	x_{{\rm fc},k+1} = A x_{{\rm da},k} + B \hat{d}_{k},	\label{kalman_fc_state}\\
	y_{{\rm fc},k} =  C x_{{\rm fc},k}, \label{kalman_fc_output}\\
	z_k = y_{{\rm fc},k} - y_k, 		\label{innov_error}
\end{gather}
where
$x_{\rm da,k} \in \mathbb R^{n}$ is the data-assimilation state, 
$x_{{\rm fc},k} \in \mathbb R^{n}$ is the forecast state,
$\hat d_k$ is the estimate of $d_k$, 
$y_{\rmf\rmc,k} \in \mathbb R$ is the forecast output,
$z_k \in \mathbb R$ is the residual, and $x_{{\rm fc},0} = 0$.

Next, to obtain $\hat{d}_k$, the input-estimation subsystem of order $n_\rme$ is given by the exactly proper, input-output dynamics
%
\begin{align}
\hat{d}_k = \sum\limits_{i=1}^{n_\rme} P_{i,k} \hat{d}_{k-i} + \sum\limits_{i=0}^{n_\rme} Q_{i,k} z_{k-i}, \label{estimate_law1}
\end{align}
%
where $P_{i,k} \in \BBR$ and $Q_{i,k} \in \BBR$.
AIE minimizes a cost function that depends on $z_{k}$ by updating $P_{i,k}$ and $Q_{i,k}$ as shown below.
The subsystem \eqref{estimate_law1} can be reformulated as
%
\begin{align}
\hat{d}_k=\Phi_k \theta_k, \label{estimate_law12}
\end{align}
where the estimated coefficient vector $\theta_k \in \mathbb{R}^{l_{\theta}}$ is defined by
%
\begin{align}
\hspace{-0.2cm}\theta_k \isdef \begin{bmatrix}
P_{1,k} & \cdots & P_{n_{\rme},k} & Q_{0,k} & \cdots & Q_{n_{\rme},k}
\end{bmatrix}^{\rm T}, \label{est_coeff_vec}
\end{align}
the regressor matrix $\Phi_k \in \mathbb{R}^{1 \times l_{\theta}}$ is defined by
%
\begin{align}
	\hspace{-0.2cm}\Phi_k \isdef
		\begin{bmatrix}
			\hat{d}_{k-1} &
			\cdots &
			\hat{d}_{k-n_{\rme}} &
			z_k &
			\cdots &
			z_{k-n_{\rme}}
		\end{bmatrix},
\end{align}
and $l_\theta \isdef  2n_{\rme} +1$.
The subsystem \eqref{estimate_law1} can be written using backward shift operator $\bfq^{-1}$ as
%
\begin{align}
   \hat{d}_{k} = G_{\hat{d}z,k}(\bfq^{-1})z_k,
\end{align}
where
%
\begin{align}
    G_{\hat{d}z,k} &\isdef D_{\hat{d}z, k}^{-1}  \it{N}_{\hat{d}z,k}, \label{d_hat_z_tf} \\
    D_{\hat{d}z,k}(\bfq^{-1}) &\isdef I_{l_d}-P_{1,k}\bfq^{-1} - \cdots-P_{n_\rme,k}\bfq^{-n_\rme}, \label{d_hat_z_tf_D} \\
    N_{\hat{d}z, k}(\bfq^{-1}) &\isdef Q_{0,k} + Q_{1,k} \bfq^{-1}+\cdots+Q_{n_\rme,k}\bfq^{-n_\rme}. \label{d_hat_z_tf_N}
\end{align}
Next, define the filtered signals
%
\begin{align}
\Phi_{{\rm f},k} &\isdef G_{{\rm f}, k}(\bfq^{-1}) \Phi_{k}, \quad
\hat{d}_{{\rm f},k} \isdef G_{{\rm f}, k}(\bfq^{-1}) \hat{d}_{k}, \label{eq:filtdhat}
\end{align}
where, for all $k\ge 0$,
%
\begin{align}
G_{{\rm f}, k}(\bfq^{-1}) = \sum\limits_{i=1}^{n_{\rm f}} \bfq^{-i}H_{i,k}, \label{Gf}
\end{align}
%
\begin{align}
H_{i,k} &\isdef \left\{
\begin{array}{ll}
C B, & k\ge i=1,\\
C \overline{A}_{k-1}\cdots \overline{A}_{k-(i-1)}  B, & k\ge i \ge 2, \\
0, & i>k,
\end{array}
\right. 
\end{align}
and $\overline{A}_k \isdef A(I + K_{{\rm da},k}C)$, where $K_{{\rm da},k}$ is the Kalman filter gain given by \eqref{kalman_gain} below.
Furthermore, for all $k \ge 0$, define the {\it retrospective performance variable} $z_{{\rm r},k} \colon \BBR^{l_\theta} \rightarrow \BBR$ by
%
\begin{align}
z_{{\rm r},k}(\hat{\theta}) \isdef z_k -( \hat{d}_{{\rm f},k} - \Phi_{{\rm f},k}\hat{\theta} ), \label{eq:RetrPerfVar} 
\end{align}
and define the \textit{retrospective cost function} $\SJ_k \colon \BBR^{l_\theta} \rightarrow \BBR$ by
%
%
\begin{align}
   \nonumber \SJ_k(\hat{\theta}) \isdef  \sum\limits_{i=0}^k  \left(\prod_{j=1}^{k-i} \lambda_{j}\right) [R_z z_{{\rm r},i}^{2}(\hat{\theta}) +  R_{\rmd} (\Phi_i\hat{\theta})^2] \\ + \left(\prod_{j=1}^{k} \lambda_{j}\right) (\hat{\theta} - \theta_0)^\rmT R_{\theta} (\hat{\theta} - \theta_0),
    \label{costf}
\end{align}
where $R_z\in(0,\infty)$, $R_d\in(0,\infty)$, $\lambda_k \in (0, 1]$ is the forgetting factor, and the regularization weighting matrix $R_{\theta}\in\BBR^{l_{\theta} \times l_{\theta}}$ is positive definite.
Then, for all $k\ge 0$, the unique global minimizer 
\begin{equation}
\theta_{k+1} \triangleq \argmin_{\hat{\theta} \in \BBR^{l_\theta}} \SJ_k(\hat{\theta}) 
\end{equation}
is given recursively by the RLS update equations \cite{islam2019recursive, lai2022exponential}
%
%
\begin{align}
P_{k+1}^{-1} &= \lambda_{k}P_k^{-1} + (1-\lambda_k)R_\infty + \widetilde{\Phi}_k^\rmT \widetilde{R} \widetilde{\Phi}_k, \label{covariance_update} \\
\theta_{k+1} &= \theta_{k} - P_{k+1} \widetilde{\Phi}^{\rm T}_{k} \widetilde{R} (\widetilde{z}_{k} + \widetilde{\Phi}_{k} \theta_{k}), \label{theta_update}
\end{align}
where $P_0 \isdef R_\theta^{-1}$, for all $k \ge 0$, $P_k \in \BBR^{l_\theta \times l_\theta}$ is the positive-definite covariance matrix, the positive-definite matrix $R_\infty \in \BBR^{l_\theta \times l_\theta}$ is the user-selected \textit{resetting matrix}, and where, for all $k \ge 0$,
%
\begin{gather*}
\widetilde{\Phi}_k \isdef \begin{bmatrix}
   \Phi_{\rmf, k}  \\
   \Phi_k   \\
\end{bmatrix}, \quad 
\widetilde{z}_k \isdef \begin{bmatrix}
   z_k-\hat{d}_{{\rm f},k}  \\
   0   \\
\end{bmatrix}, \quad
\widetilde{R} \isdef \begin{bmatrix}
   R_z & 0  \\
   0 & R_{\rmd}   \\
\end{bmatrix}.
\end{gather*}
Hence, \eqref{covariance_update} and \eqref{theta_update} recursively update the estimated coefficient vector \eqref{est_coeff_vec}.

The forgetting factor $\lambda_k \in(0,1]$ in \eqref{costf} and \eqref{covariance_update} enables the eigenvalues of $P_k$ to increase, 
which facilitates  adaptation of the input-estimation subsystem \eqref{estimate_law1}  
\cite{aastrom1977theory}.
In addition, the resetting matrix $R_\infty$ in \eqref{covariance_update} prevents the eigenvalues of $P_k$ from becoming excessively large under conditions of poor excitation \cite{lai2022exponential}, a phenomenon known as covariance windup \cite{malik1991some}.
{Variable-rate forgetting based on the \textit{F}-test is used to select the forgetting factor $\lambda_k \in (0,1]$. Additional details are given in \cite{mohseni2022recursive, verma_shashank_2024_realtime_VRF_axiv}.}

\subsection{State Estimation}

The forecast variable $x_{{\rm fc},k}$ updated by \eqref{kalman_fc_state} is used to obtain the estimate $x_{{\rm da},k}$ of $x_k$ given, for all $k \ge 0$, by the Kalman filter data-assimilation step
\begin{align}
x_{{\rm da},k} &= x_{{\rm fc},k} + K_{{\rm da},k} z_k, \label{kalman_da_state}
\end{align}
where the Kalman filter gain $K_{{\rm da},k} \in \mathbb R^{n}$, the data-assimilation error covariance $P_{{\rm da},k} \in \mathbb R^{n \times n},$
and the forecast error covariance $P_{\rmf\rmc,k+1} \in \mathbb R^{n \times n}$ are given by
\begin{align}
    K_{{\rm da},k} &= - P_{\rmf\rmc,k}C^{\rm T} ( C P_{\rmf\rmc,k} C^{\rm T} + V_{2,k}) ^{-1}, \label{kalman_gain} \\
    P_{{\rm da},k} &=  (I_{n}+K_{{\rm da},k}C) P_{\rmf\rmc,k},\label{Pda} \\
	P_{\rmf\rmc,k+1} &=  A P_{{\rm da},k}A^{\rm T} + V_{1,k}, \label{Pf}
\end{align}
where $V_{2,k} \in \mathbb R$ is the measurement noise covariance, $V_{1,k}$ is defined by 
\begin{align}
    \nonumber V_{1,k}\isdef\,  B {\rm var}(d_k-\hat{d}_k)B^\rmT 
     \\ \nonumber + A {\rm cov}(x_k - x_{{\rm da},k},d_k-\hat{d}_k)B^\rmT 
     \\+ B {\rm cov}(d_k-\hat{d}_k,x_k - x_{{\rm da},k})A^\rmT,
\end{align}
and $P_{\rmf\rmc,0} = 0.$ 
\subsection{Adaptive State Estimation} \label{sec:AdapInptStateEst}

This section summarizes the adaptive state estimation component of AISE. 
Assuming that, for all $k \ge 0$, $V_{1,k}$ and $V_{2,k}$ are unknown in (\ref{Pf}) and \eqref{kalman_gain},
the goal is to adapt ${V}_{{1,\rm adapt},k}$ and ${V}_{{2,\rm adapt},k}$ at each step $k$ to estimate $V_{1,k}$ and $V_{2,k}$.
To do this, we define, for all $k \ge 0$, the 
performance metric $J_k \colon \BBR^{n\times n} \times \BBR \rightarrow \BBR$ by
\begin{align}
  {J}_{k}({V}_{1},{V}_{2}) \isdef |\widehat{S}_{ k}-{S}_{ k}|, \label{J_daptmetric}
\end{align}
where $\widehat{S}_{ k}$ is the sample variance of $z_k$ over $[0,k]$ defined by
\begin{align}
    \widehat{S}_{k} \isdef \cfrac{1}{k}\sum^{k}_{i=0}(z_i - \overline{z}_k)^2, \quad
    \overline{z}_k \isdef \cfrac{1}{k+1}\sum^{k}_{i=0}z_i,  \label{var_comp}
\end{align}
and ${S}_{k}$ is the variance of the residual $z_k$ determined by the Kalman filter, given by
\begin{align}
    {S}_{k} \isdef  C (A P_{{\rm da},k-1}A^{\rm T} + V_{1}) C^{\rm T} + V_{2}.  \label{var_inno}
\end{align}
For all $k \ge 0$, we assume for simplicity that 
\begin{align}
{V}_{{1,\rm adapt},k}  \triangleq \eta_k I_n, \label{eq:V1adapt_k}
\end{align}
and we define the set $\SSS$ of minimizers 
$(\eta_k,{V}_{{2,\rm adapt},k})$ of $J_k$ by
\begin{align}
      \SSS\isdef \{ (\eta_k,{V}_{{2,\rm adapt},k}) \colon \eta \in [\eta_{\rmL},\eta_{\rmU}] \mbox{ and }\\ {V}_{2} \ge 0 \mbox{ minimize }  J_k(\eta I_{n},V_{2})\}, \label{covmin}
\end{align} 
where
$0 \le \eta_{\rmL} \le \eta_{\rmU}.$
Next, defining ${J}_{\rmf,k} \colon \BBR \rightarrow \BBR $ by
\begin{align}
    {J}_{\rmf,k}(V_{1}) \isdef \widehat{S}_{k} - C (A P_{{\rm da},k-1}A^{\rm T} + V_{1})  C^{\rm T}, \label{J1_func}
\end{align}
and using \eqref{var_inno}, it follows that (\ref{J_daptmetric}) can be written as
\begin{align}
    {J}_k({V}_{1},{V}_{2}) = |{J}_{\rmf,k}(V_{1})-V_{2}|. \label{J_daptmetric_V2}
\end{align}
We then construct the set $\SJ_{\rmf,k}$ of positive values of ${J}_{\rmf,k}$ given by 
\begin{align}
      \SJ_{\rmf,k} \isdef \{J_{\rmf,k}(\eta I_{n}) \colon J_{\rmf,k}(\eta I_{n}) > 0, \eta_{\rmL} \le\eta \le\eta_{\rmU}\} \subseteq \BBR. \label{J_f_positive}
\end{align}
Following result provides a technique for computing $\eta_k$ and ${V}_{{2,\rm adapt},k}$ defined in \eqref{covmin}.

\begin{prop}\label{prop: eta_k and V2,adapt minimizer}
    Let $k \ge 0$.  Then, the following statements hold:
    \begin{enumerate}
        \item  Assume that $\SJ_{\rmf,k}$ is nonempty, let $\beta \in [0,1]$, and define $\eta_k$ and $V_{2,k}$ by
    \begin{gather}
        \eta_k = \underset{\eta \in [\eta_L,\eta_U]}{\arg \min} \ |J_{\rmf,k}(\eta I_{n}) -  \widehat{J}_{\rmf,k}(\beta)|, \label{eq:eta_opt_1}\\
        {V}_{{2,\rm adapt},k} = J_{\rmf,k}(\eta_k I_n),
        \label{v_2_opt_1_NE}
    \end{gather}
    where
\begin{align}
       \widehat{J}_{\rmf,k}(\beta) \isdef \beta \min \SJ_{\rmf,k}+(1-\beta)\max \SJ_{\rmf,k}. \label{alpha1}
\end{align}
Then, $(\eta_k,{V}_{{2,\rm adapt},k})\in\SSS.$
\item      Assume that $\SJ_{\rmf,k}$ is empty, and define $\eta_k$ and $V_{2,k}$ by
\begin{gather}
        \eta_k = \underset{\eta \in [\eta_\rmL,\eta_\rmU]}{\arg \min} \ |J_{\rmf,k}(\eta I_{n})|, \label{eq:eta_opt_2}\\
        {V}_{{2,\rm adapt},k} = 0. \label{v_2_opt_1_E}
\end{gather}
Then, $(\eta_k,{V}_{{2,\rm adapt},k})\in\SSS.$
    \end{enumerate}
\end{prop}

\textit{Proof:} See Section 5.2 of \cite{verma_shashank_2023_realtime_IJC}. \hfill $\square$

\begin{figure}[h!t]
  \begin{center}
{\includegraphics[width=0.9\linewidth]{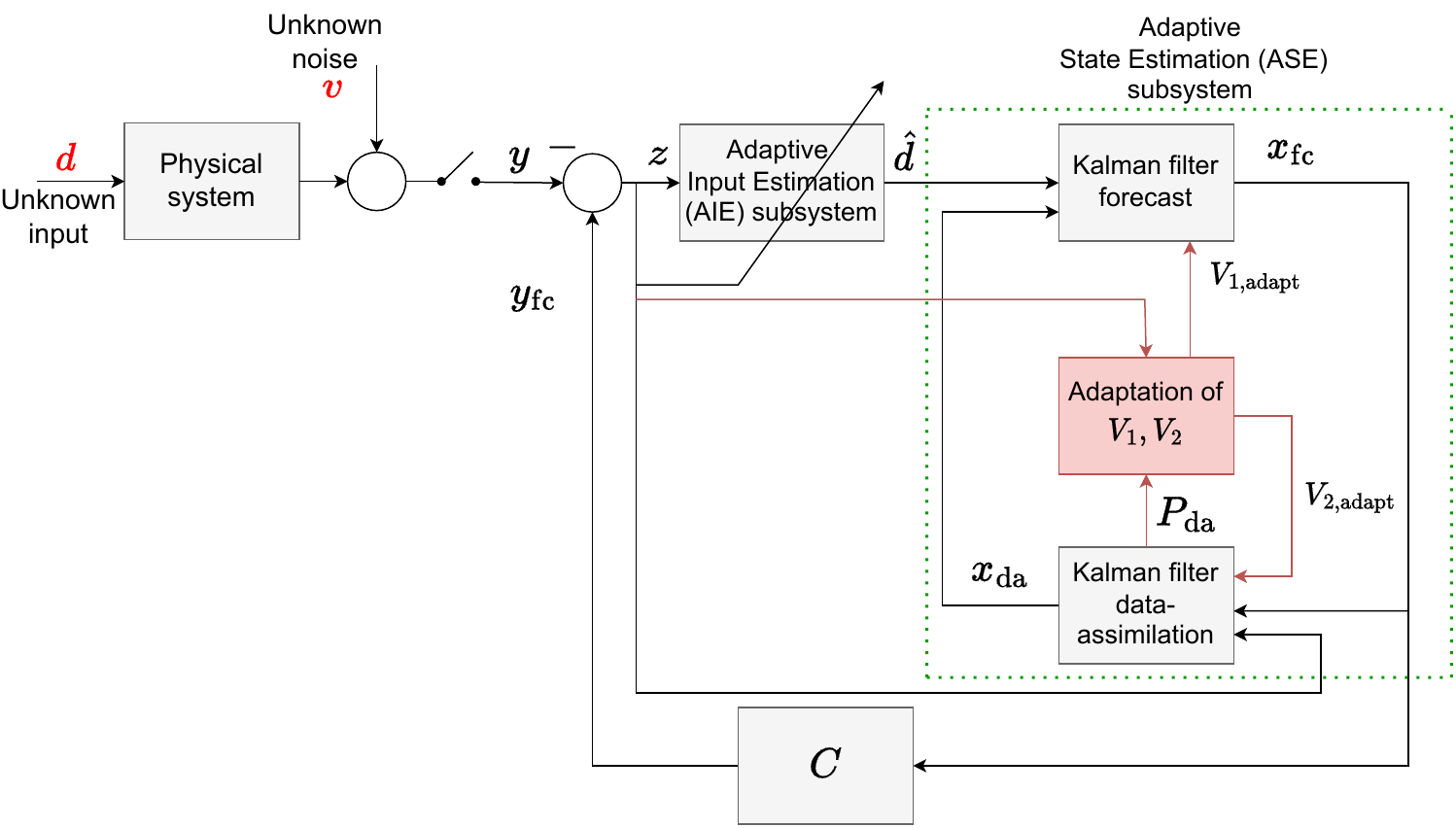}}
\end{center}
\caption{Block diagram of AISE.}
\label{fig:sec5_block_diag_AIE_ASE}
\end{figure}

A block diagram of AISE is shown in Figure \ref{fig:sec5_block_diag_AIE_ASE}. 
Hence, at each step $k\ge0,$ $\hat{d}_k$ is computed from the input $y_k,$ such that
\begin{equation}
    \hat{d}_k = f_{{\rm aise},k} (y_k),
\end{equation}
where $f_{{\rm aise},k} \colon \BBR \to \BBR$ encodes the operations performed by \eqref{kalman_fc_state}$-$\eqref{estimate_law12},
\eqref{eq:filtdhat},
\eqref{Gf},
\eqref{covariance_update},
\eqref{theta_update},
\eqref{kalman_da_state}$-$\eqref{Pf},
\eqref{eq:V1adapt_k},
\eqref{eq:eta_opt_1},
\eqref{v_2_opt_1_NE},
\eqref{eq:eta_opt_2},
\eqref{v_2_opt_1_E}.
Note that $f_{{\rm aise},k}$ depends on the current step $k$ since several internal variables are updated at each step.

\section{Trajectory Prediction using AISE/va} \label{sec:aise_va}
At step $k$,  to predict the future trajectory $\hat{p}_{k+l}$ for $l \in \{1,\ldots,\ell\}$, we consider AISE/va.  
%
AISE/va uses the velocity and acceleration in \eqref{vel_resolve_E} and \eqref{acc_resolve_E} with the second-order approximation 
\begin{align}
    \hat{p}_{k+l} &\isdef p_{k} + l T_{\rms}\hat{{v}}_{k} + \tfrac{1}{2} l^2T_{\rms}^2 \hat{{a}}_{k}, \label{pred_taylor}
\end{align}
where $\hat{{v}}_{k}$ and $\hat{{a}}_{k}$ are estimates of ${{v}}_{k}$ and ${{a}}_{k}$,
$k+l$ is a future time step, and $\ell$ is the horizon.
Note that \eqref{pred_taylor} assumes that the velocity and acceleration are constant over the horizon. 
The derivative estimates $\hat{{v}}_{k}$ and $\hat{{a}}_{k}$ in \eqref{pred_taylor} are computed using AISE, which is summarized in Section \ref{sec:AISE}. AISE/va uses the estimates of velocity and acceleration obtained using AISE for trajectory prediction with horizon $\ell$ steps.

\section{Trajectory Prediction using AISE/FS} \label{sec:aise_fs}
At step $k$,  to predict the future trajectory $\hat{p}_{k+l}$ for $l \in \{1,\ldots,\ell\}$, we consider AISE/FS.  
Since the target moves in 3D space along the trajectory $p(t) \in \mathbb{R}^3$, it follows that
\begin{align}
    \dot{p}(t) = u(t) T(t), \label{FS_pos}
\end{align}
where $T(t)$ is the unit tangent vector and $u(t) \geq 0$ is the speed along $T(t)$.
The Frenet-Serret formulas are given by
\begin{align}
    \dot{T}(t) &= u(t) \Tilde{\kappa}(t) N(t), \label{FS_eq_T} \\
    \dot{N}(t) &= u(t) [-\Tilde{\kappa}(t) T(t) + \Tilde{\tau}(t) B(t)], \label{FS_eq_N}  \\
    \dot{B}(t) &= -u(t)\Tilde{\tau}(t) N(t), \label{FS_eq_B}  
\end{align} 
where $\Tilde{\kappa}(t)$ is the curvature, $\Tilde{\tau}(t)$ is the torsion, and $N(t)$ is the unit normal vector and $B(t)$ is the unit binormal vectors. 
The vectors $T(t)$, $N(t),$ and $B(t)$ are mutually orthogonal. For brevity, the time variable $t$ will be omitted henceforth.

The Frenet-Serret formulas \eqref{FS_eq_T}, \eqref{FS_eq_N}, and \eqref{FS_eq_B},  can be written as \cite{bonnabel_cdc_2017_target_tracking}
\begin{align}
     \begin{bmatrix}
    \dot{T} & \dot{N} & \dot{B} 
    \end{bmatrix} = u \begin{bmatrix}
    T & N & B 
    \end{bmatrix} \begin{bmatrix}
    0 & -\Tilde{\kappa} & 0 \\ 
    \Tilde{\kappa} & 0 & -\Tilde{\tau}  \\ 
    0 & \Tilde{\tau} & 0
    \end{bmatrix} . \label{FS_eq_matrix}
\end{align} 
Defining $\kappa \isdef u\Tilde{\kappa}$ and $\tau \isdef u\Tilde{\tau}$, we write \eqref{FS_eq_matrix} as
\begin{align}
    \begin{bmatrix}
    \dot{T} & \dot{N} & \dot{B} 
    \end{bmatrix} =  \begin{bmatrix}
    T & N & B 
    \end{bmatrix} \begin{bmatrix}
    0 & -{\kappa} & 0 \\ 
    {\kappa} & 0 & -{\tau}  \\ 
    0 & {\tau} & 0
    \end{bmatrix} . \label{FS_eq_matrix_rewrite}
\end{align} 
Defining the orthogonal matrix $R \isdef \begin{bmatrix}
    T & N & B 
    \end{bmatrix}$, \eqref{FS_eq_matrix_rewrite} and   \eqref{FS_pos} can be written as
\begin{align}
     \dot{R} &= R\omega^{\times}, \label{rot_mat_conti} \\
     \dot{p} &= R \begin{bmatrix} u & 0 & 0 \end{bmatrix}^\rmT,\label{pos_pred_conti}
\end{align} 
where $(\cdot)^{\times}$ denotes a $3 \times 3$ skew-symmetric matrix and $\omega \isdef \begin{bmatrix} \tau & 0 & \kappa \end{bmatrix}^\rmT$.

Using the position, velocity, and acceleration defined by \eqref{pos_resolve_E}, \eqref{vel_resolve_E}, and \eqref{acc_resolve_E}, the tangent, normal, and binormal vectors at step $k$ are given by (excluding nongeneric cases)  \cite{Hanson_FS_formula_1994} 
\begin{align}
    T_k &= \frac{p_k}{\Vert 
p_k \Vert}, \label{T_k}\\
N_k &= \frac{v_k \times (a_k \times v_k)}{\Vert 
v_k \Vert  \Vert a_k \times v_k \Vert}, \label{N_k}\\
B_k &= \frac{v_k \times a_k}{\Vert 
v_k \times a_k \Vert}, \label{B_k}
\end{align} 
where $T_k \isdef T(kT_\rms)$, $N_k \isdef N(kT_\rms),$ and $B_k \isdef B(kT_\rms).$
Similarly, $u$, $\Tilde{\kappa}$, and $\Tilde{\tau}$ at step $k$ are given by \cite{Hanson_FS_formula_1994} 
\begin{align}
    u_k &= \Vert 
v_k \Vert, \label{u_norm}\\
    \Tilde{\kappa}_k &= \frac{\Vert v_k \times a_k \Vert}{\Vert 
v_k \Vert^3}, \label{curvature}\\
\Tilde{\tau}_k &= \frac{v_k^\rmT  (a_k \times j_k)}{\Vert v_k \times a_k
\Vert^2}, \label{torsion}
\end{align}
where $u_k \isdef u(kT_\rms)$, $\Tilde{\kappa}_k \isdef \Tilde{\kappa}(kT_\rms),$  $\Tilde{\tau}_k \isdef \Tilde{\tau}(kT_\rms),$ and the jerk is defined by
\begin{align}
j_k \isdef  \dddot{p}_k 
 = \, \framedddotE{\vect{r}}_{\rmo_{\rm T}/\rmo_{\rmE}}(t_k)\bigg\vert_\rmE \in \mathbb{R}^3.  \label{jerk_resolve_E}
\end{align}
%
%

Assuming zero-order hold, integrating \eqref{rot_mat_conti} from  $kT_\rms$ to $(k+1)T_\rms$ yields \cite{hartley_contact_aided_2020}
\begin{align}
     R_{k+1} = R_k \text{exp}(\omega_k^{\times}T_\rms), \label{R_update_exp}
\end{align}
where 
\begin{align}
\omega_k =  \begin{bmatrix} \tau_k & 0 & \kappa_k \end{bmatrix}^\rmT =  \begin{bmatrix} u_k\Tilde{\tau}_k & 0 & u_k\Tilde{\kappa}_k \end{bmatrix}^\rmT \label{omega_k}.
\end{align}
Likewise, integrating \eqref{pos_pred_conti} from  $kT_\rms$ to $(k+1)T_\rms$ yields \cite{hartley_contact_aided_2020}
\begin{align}
     p_{k+1} = p_k + R_k \bigg( \int_{kT_\rms}^{(k+1)T_\rms} \text{exp}(\omega_k^{\times}t)\,\rmd t\bigg)\begin{bmatrix} u_k & 0 & 0 \end{bmatrix}^\rmT. \label{pos_update_exp}
\end{align}
Next, defining \cite{Barfoot_2017,Chirikjian_2017}
\begin{align}
     \Gamma_{0}(\phi) \isdef I_{3} + \frac{\sin(\Vert\phi\Vert)}{\Vert\phi\Vert}\phi^{\times} + \frac{1 - \cos(\Vert\phi\Vert)}{\Vert\phi\Vert^2}\phi^{\times 2}, \label{gamma_0}
\end{align}
\begin{align}
     \Gamma_{1}(\phi) \isdef I_{3} + \frac{1 - \cos(\Vert\phi\Vert)}{\Vert\phi\Vert^2}\phi^{\times} +   \frac{\Vert\phi\Vert - \sin(\Vert\phi\Vert)}{\Vert\phi\Vert^3}\phi^{\times 2}, \label{gamma_1}
\end{align}
%
%
it follows that
\begin{align}
\text{exp}(\omega_k^{\times}T_\rms) = \Gamma_0(\omega_k T_\rms),
\end{align}
\begin{align}
\int_{kT_\rms}^{(k+1)T_\rms} \text{exp}(\omega_k^{\times}t)\, \rmd t = \Gamma_{1}(\omega_k T_\rms)T_\rms.
\end{align}
Using \eqref{gamma_0} and \eqref{gamma_1}, we write \eqref{R_update_exp} and \eqref{pos_update_exp} as the discrete-time dynamics 
\begin{align}
     R_{k+1} &= R_k \Gamma_0(\omega_k T_\rms), \label{R_update_k} \\
     p_{k+1} &= p_k + T_\rms R_k \Gamma_1(\omega_k T_\rms)  \begin{bmatrix} u_{k} & 0 & 0 \end{bmatrix}^\rmT. \label{pos_update_k}
\end{align}
%
%
%
To predict the trajectory $l$ steps into the future, it follows from \eqref{R_update_k} and \eqref{pos_update_k} that, for all $l \in \{1, \ldots, \ell\}$,
\begin{align}
     \hat{R}_{k+l} &= R_k [ \Gamma_0(\omega_k T_\rms) ]^l, \label{R_update_l} \\
     \hat{p}_{k+l} &= p_k + T_\rms   \bigg[R_k + \sum_{i=1}^{l-1} \hat{R}_{k+i}\bigg]\Gamma_1(\omega_k T_\rms) \begin{bmatrix} u_{k} & 0 & 0 \end{bmatrix}^\rmT.\label{pos_update_l}
\end{align}
Note that \eqref{R_update_l} and \eqref{pos_update_l} assume that, for all $l \in \{1, \ldots, \ell\}$, $\omega_{k+l} = \omega_k$ and $u_{k+l}=u_k.$

Using numerical differentiation, AISE/FS uses the position measurement $p_k$ to compute the velocity estimate $\hat{v}_k$, the acceleration estimate $\hat{a}_k$, and the jerk estimate $\hat{j}_k$. 
These estimates are used to compute the Frenet-Serret parameters \eqref{u_norm}, \eqref{curvature}, \eqref{torsion}, and \eqref{omega_k}, which approximate the 3D curve followed by the target.
These parameters are used by \eqref{pos_update_l} for trajectory prediction with horizon $l = \ell$ steps.
%
\subsection{Summary}
\begin{figure}[h!t]
              \begin{center}
            {\includegraphics[width=1.08\linewidth]{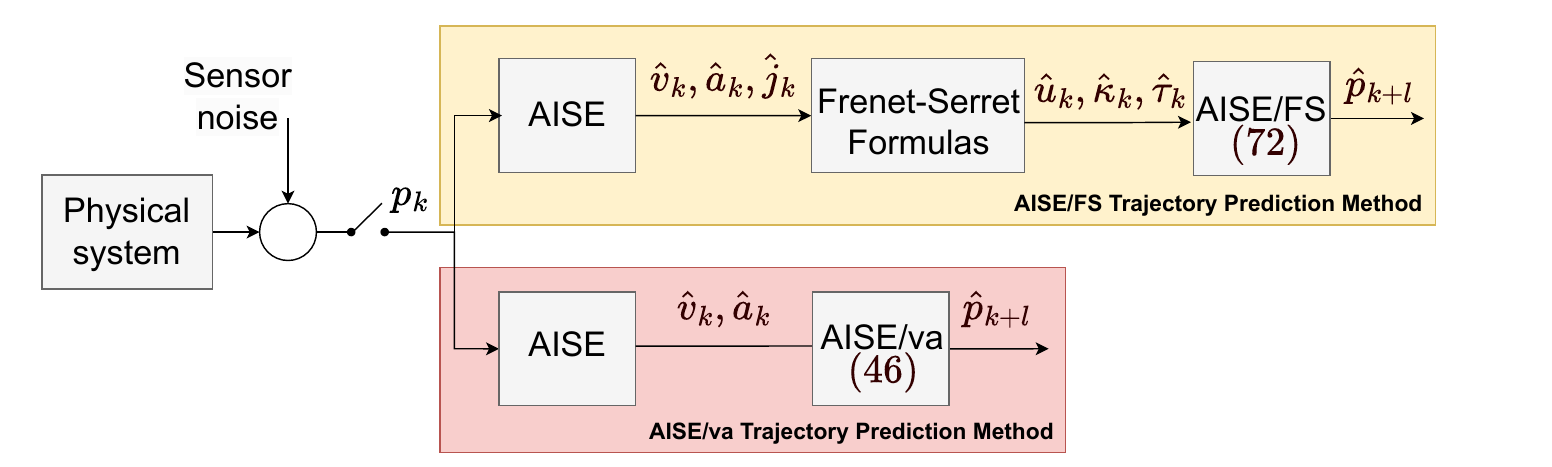}}
            \end{center}
            \caption{Block diagram of AISE/va and AISE/FS trajectory prediction methods.} 
            \label{fig:block_giag_AISE_pred}
          \end{figure}

Table \ref{Tab:pred_methods} summarizes the trajectory prediction given by \eqref{pred_taylor} and \eqref{pos_update_l} with $v_k$, $a_k$, and $j_k$ replaced by the estimates $\hat{v}_k$, $\hat{a}_k$, and $\hat{j}_k$ obtained from AISE. 
Additionally, Figure \ref{fig:block_giag_AISE_pred} shows the block diagram of AISE/va and AISE/FS.
\begin{table}[h!t]
\begin{center} 
\begin{tabular}{ |c|c|c|c| }
 \hline
 \textbf{\makecell{Trajectory Prediction \\ Method}} & \textbf{\makecell{Prediction \\ Equation}} &  {\textbf{\makecell{Estimates from \\ AISE} }}  \\
 \hline
  AISE/va & \eqref{pred_taylor}  & \makecell{$\hat{v}_k$, $\hat{a}_k$}  \\
  \hline
  AISE/FS & \eqref{pos_update_l}  & \makecell{$\hat{v}_k$, $\hat{a}_k$, $\hat{j}_k$}  \\
  \hline
\end{tabular}
\end{center}
\captionof{table}{AISE uses the position data ${p}_{k}$ to compute the estimates $\hat{v}_k$, $\hat{a}_k$, and $\hat{j}_k$. AISE/va and AISE/FS use these estimates to predict the trajectory with horizon $\ell$ steps.
}\label{Tab:pred_methods}
\end{table}

%

 

\section{Numerical Examples} \label{sec:num_example}
In this section, two numerical examples are used to compare AISE/va and AISE/FS. For further comparison, velocity and acceleration estimates $\hat{v}_k$ and $\hat{a}_k$ in \eqref{pred_taylor} are also obtained from two additional numerical differentiation methods, namely, the backward difference with Butterworth filter (BDB) and the $\alpha$-$\beta$-$\gamma$ filter (ABG). In BDB, noisy position measurements are first smoothed using a Butterworth filter, followed by the backward difference. The tracking index $\Gamma$ is a key parameter in the $\alpha$-$\beta$-$\gamma$ filter \cite{Kalata1983TheTI}. Table \ref{table_vel_acc} summarizes the prediction methods.

\begin{table}[h!t]
\begin{center} 
\begin{tabular}{ |c|c|c|c| }
 \hline
 Prediction Methods & $\hat{{v}}_{k}$ & $\hat{{a}}_{k}$ & $\hat{{j}}_{k}$ \\
 \hline
 BDB/va   & Used  & {Used} & {Not used} \\
 \hline
 ABG/va  & Used  & Used & {Not used} \\
 \hline 
 AISE/va   & Used  & {Used} & {Not used}\\
 \hline
 AISE/FS   & Used  & Used & Used \\
 \hline
\end{tabular}
\end{center}
\caption{Estimates used in the prediction methods BDB/va, ABG/va, and AISE/va.  AISE/FS requires estimates of the first, second, and third derivatives  of ${{p}}_{k}$. }
\label{table_vel_acc}
\end{table}

To assess the accuracy of the predicted trajectory, we define the root-mean-square error (RMSE) metric with  horizon $\ell$ steps
\begin{align}
 {\rm RMSE}_{\ell} \isdef
\frac{1}{\widetilde{N}}\sqrt{{\sum_{k=k_0}^{N-\ell}({p}_{k+\ell}-\hat{p}_{k+\ell})^2}} \in \mathbb{R}^{3},  \label{rms}
\end{align}
where $\widetilde{N} = N-\ell - (k_0 -1)$.  To avoid the transient adaptation of AISE, $k$ starts from $k_0 = 2000$ in \eqref{rms}. Note that the three-axis components of \eqref{rms}, defined as ${\rm RMSE}_{x,\ell}$, ${\rm RMSE}_{y,\ell}$, and ${\rm RMSE}_{z,\ell} \in \mathbb{R}$, represent the RMSE values in the $x$, $y$, and $z$ directions.

\begin{exam} \label{eg:traj_extra_parabola}
      {\it Trajectory Prediction for a Parabolic Trajectory.}
      {\rm In this scenario, the target follows a parabolic trajectory in the $x$-$y$ plane with uniform gravity $9.8$ m/s$^2$ in the negative $y$ direction. The discrete-time position is given by  
      \begin{align}
      p_k = \begin{bmatrix}
           400 k T_{\rms} & 400 k T_{\rms} - 9.8 \frac{(kT_\rms)^2}{2}
      \end{bmatrix}^{\rmT}, \label{parabolic_traj}
    \end{align}
    where $T_{\rms} = 0.01$ s and $k \geq 0.$ To simulate noisy measurements of the target position, white Gaussian noise is added to $p_{k}$, with standard deviation $\sigma = 1.0$ m.  The horizon is $\ell = 100$ steps, which corresponds to 1 s.

    For single differentiation using AISE, we set $n_\rme = 25$, $n_\rmf = 50$, $R_z = 1$, $R_d = 10^{-1}$, $R_\theta = 10^{-3.5}I_{51}$, $\eta = 0.002,$ $\tau_n = 5,$ $\tau_d = 25,$ $\alpha = 0.002$, and $R_{\infty} = 10^{-4}.$ 
    The parameters $V_{1,k}$ and $V_{2,k}$ are adapted, with $\eta_{L} = 10^{-6}$, $\eta_{\rmU} = 0.1$, and $\beta = 0.55$ as described in Section \ref{sec:AdapInptStateEst}. For double differentiation using AISE, all parameters are the same as for single differentiation. For triple differentiation using AISE, the parameters are the same as for single differentiation, except $R_\theta = 10^{-6}I_{51}$ and $\beta = 0.5$. For BDB, the Butterworth filter is $10^{th}$ order with a cutoff frequency of $0.8\pi$ rad/step. For ABG,  the tracking index is $\Gamma = 0.6$. 

    Table \ref{table:rmse_values_parabolic} presents the RMSE values \eqref{rms} in the $x$ and $y$ directions, with horizon $\ell = 100$ steps, for BDB/va, ABG/va, AISE/va, and AISE/FS. Among these, AISE/FS achieves the lowest overall RMSE. Figure \ref{fig:exp_parabola_AISE_FS_2d} shows the predicted trajectory at each step for the horizon $\ell = 100$ steps. Figure \ref{fig:exp_parabola_AISE_FS_parameter} shows the estimated parameters of the Frenet-Serret frame using AISE. The estimated parameters closely match the true values.  \hfill $\diamond$

\begin{table}[h!t]
\begin{center} 
\begin{tabular}{|c|c|c|c|c|c|c|}
\hline
\textbf{Prediction Method} & {${\rm RMSE}_{x,100}$} & {${\rm RMSE}_{y,100}$}  \\
\hline
BDB/va & 396.24 & 405.86  \\
\hline
ABG/va & 4709.44 & 4324.09  \\
\hline
AISE/va & 34.90 & 32.07  \\
\hline
AISE/FS & {3.08} & {4.81}  \\
\hline
\end{tabular}
\end{center} 
\caption{RMSE values in $x$ and $y$ direction with horizon $\ell = 100$ steps for the parabolic trajectory.}
\label{table:rmse_values_parabolic}
\end{table}
 \begin{figure}[h!t]
              \begin{center}
            {\includegraphics[width=0.8\linewidth]{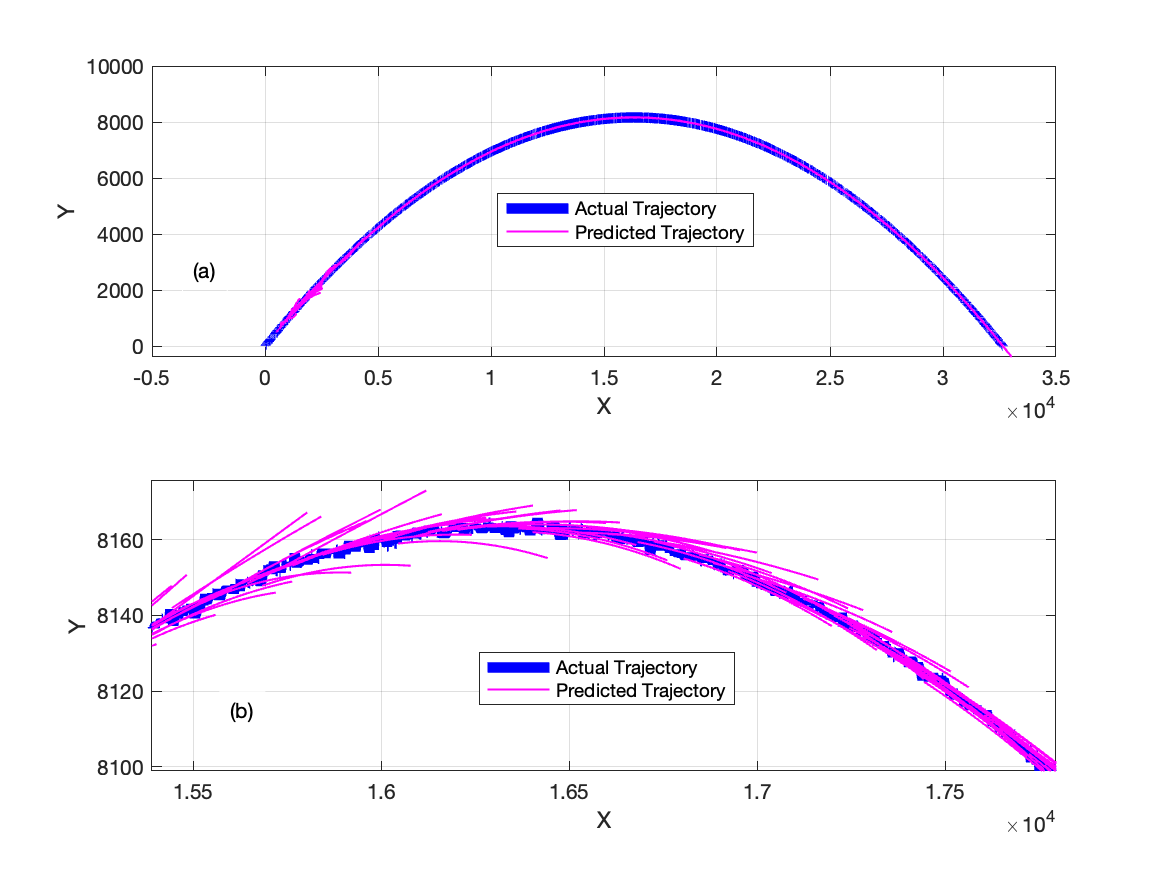}}
            \end{center}
            \caption{ {\it  Example \ref{eg:traj_extra_parabola}: Trajectory prediction for a parabolic trajectory using AISE/FS.} (a) The purple line shows the predicted trajectory with horizon $\ell = 100$ steps. (b) Zoom of (a).} 
            \label{fig:exp_parabola_AISE_FS_2d}
          \end{figure}

 \begin{figure}[h!t]
              \begin{center}
            {\includegraphics[width=1\linewidth]{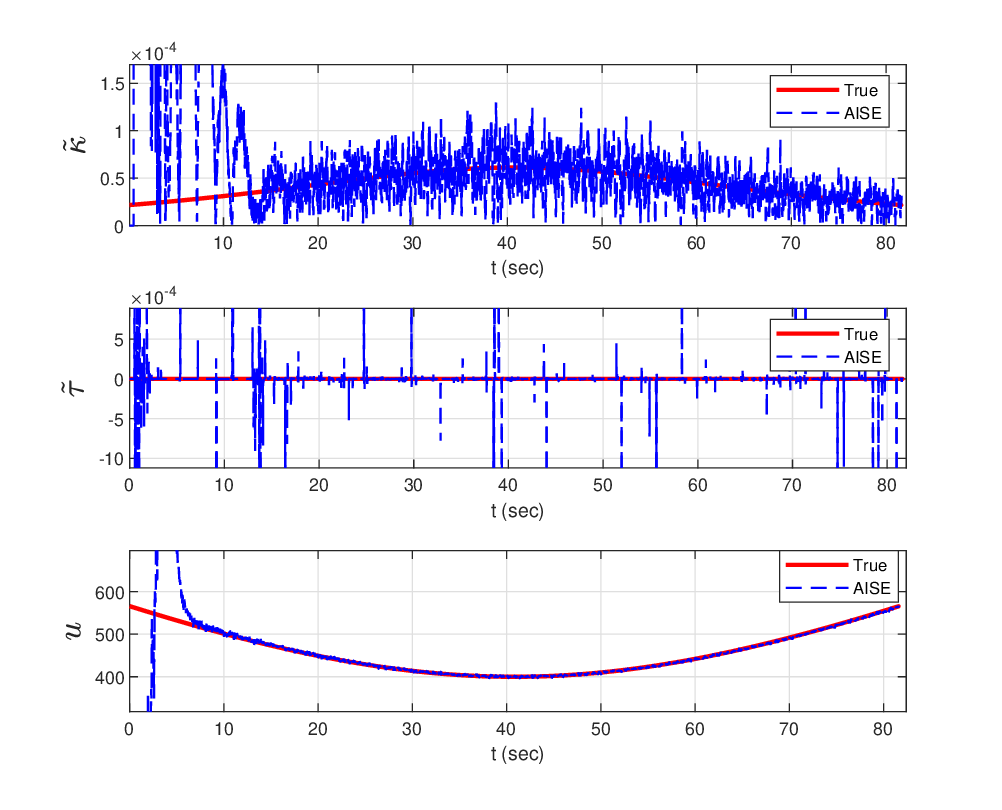}}
            \end{center}
            \caption{ {\it  Example \ref{eg:traj_extra_parabola}: Trajectory prediction for a parabolic trajectory using AISE/FS.} Estimates of $\Tilde{\kappa},$ $\Tilde{\tau}$, and $u$ are computed using AISE.} 
            \label{fig:exp_parabola_AISE_FS_parameter}
          \end{figure}

}      
\end{exam}

\begin{exam} \label{eg:traj_helix}
      {\it Trajectory Prediction for a Helical Trajectory.}
      {\rm In this scenario, the target follows a helical trajectory with discrete-time position given by 
       \begin{align}
      p_k = \begin{bmatrix}
           20 \sin (0.5 k T_{\rms}) & 20 \cos (0.5 k T_{\rms}) & k T_{\rms}
      \end{bmatrix}^{\rmT} \label{helix_traj}
    \end{align}
    %
    where $T_{\rms} = 0.01$ s and $k \geq 0.$ To simulate noisy measurements, white Gaussian noise with standard deviation $\sigma = 0.1$ m is added to each position measurement. The parameters of AISE, BDB, and ABG are the same as in Example \ref{eg:traj_extra_parabola}. 

    Table \ref{table:rmse_values_helix} presents the RMSE values \eqref{rms} in the $x,$ $y,$ and $z$ directions, with horizon $\ell = 100$ steps, for BDB/va, ABG/va, AISE/va, and AISE/FS. Among these, AISE/FS achieves the lowest overall RMSE. Figure \ref{fig:exp_helix_AISE_FS_2d} shows the predicted trajectory at each step for the horizon $\ell = 100$ steps. Figure \ref{fig:exp_helix_AISE_FS_parameter} shows the estimated parameters of Frenet-Serret using AISE. The estimated parameters closely match the true values. \hfill $\diamond$

\begin{table}[h!t]
\begin{center} 
\begin{tabular}{|c|c|c|c|c|c|c|}
\hline
\textbf{Prediction Method} & {${\rm RMSE}_{x,100}$} & {${\rm RMSE}_{y,100}$} & {${\rm RMSE}_{z,100}$} \\
\hline
BDB/va & 83.97 & 86.42 & 83.91 \\
\hline
ABG/va & 47.11 &  41.08 & 43.38  \\
\hline
AISE/va & 1.45 & 0.89 & 0.08  \\
\hline
AISE/FS & {0.46} & {0.27} & {0.05} \\
\hline
\end{tabular}
\end{center} 
\caption{RMSE values with horizon $\ell = 100$ steps   for the helical trajectory. Note that AISE/FS provides the minimum RMSE along each axis.}
\label{table:rmse_values_helix}
\end{table}

 \begin{figure}[h!t]
              \begin{center}
            {\includegraphics[width=0.8\linewidth]{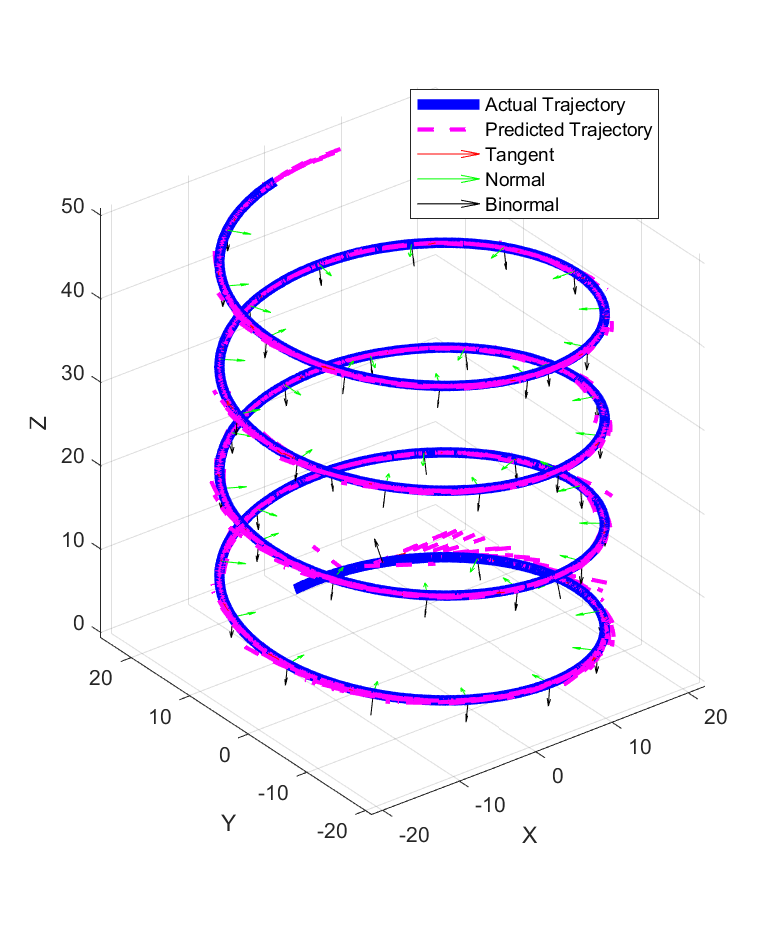}}
            \end{center}

            \caption{ {\it  Example \ref{eg:traj_helix}: Trajectory prediction for the helical trajectory using AISE/FS.} The purple line shows that the trajectory prediction with horizon $\ell = 100$ steps is close to the true trajectory.} 
            %
            \label{fig:exp_helix_AISE_FS_2d}
          \end{figure}

 \begin{figure}[h!t]
              \begin{center}
            {\includegraphics[width=1\linewidth]{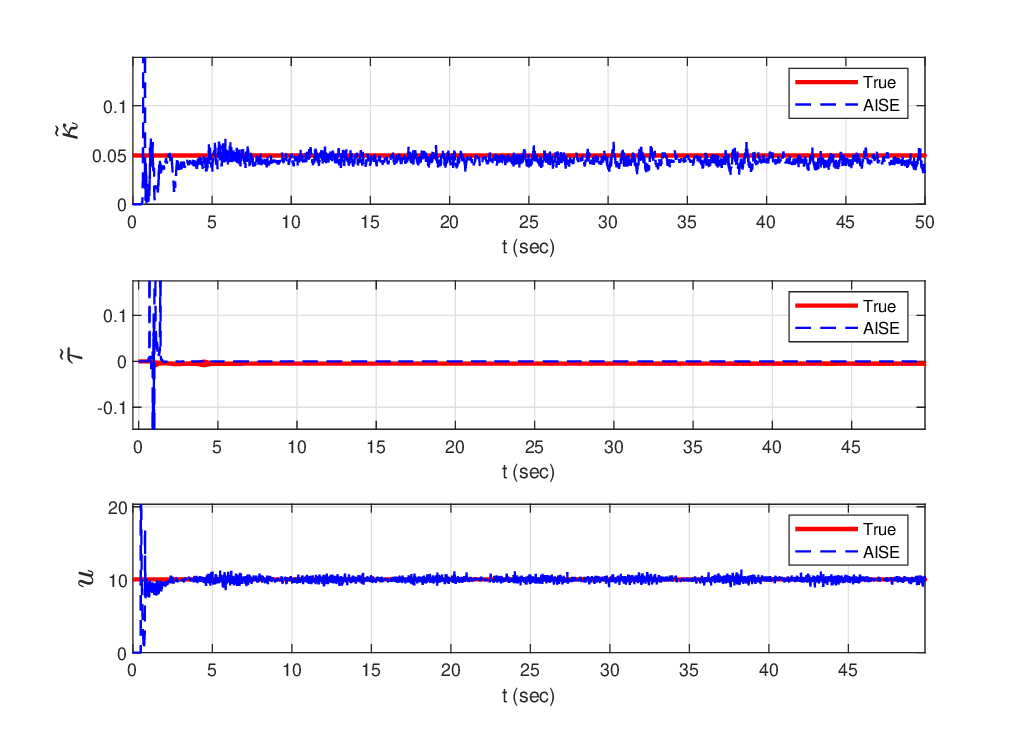}}
            \end{center}
            \caption{ {\it  Example \ref{eg:traj_helix}: Trajectory prediction for a helical trajectory using AISE/FS.} 
            Estimates of $\Tilde{\kappa},$ $\Tilde{\tau}$, and $u$ are computed using AISE.} 
            \label{fig:exp_helix_AISE_FS_parameter}
          \end{figure}
     
}      
\end{exam}

\section{CONCLUSIONS}
This paper introduced two methods, AISE/va and AISE/FS, for real-time trajectory prediction based on position measurements. 
AISE/va uses estimates of velocity and acceleration to predict the future trajectory. 
AISE/FS uses position measurements to estimate the Frenet-Serret frame of the target trajectory, which requires estimates of velocity, acceleration, and jerk to predict the future trajectory.

For both methods, adaptive input and state estimation (AISE) was used to estimate the required derivatives.
The performance of both methods was compared numerically with traditional approaches for a parabolic trajectory and a helical trajectory. AISE/FS demonstrated more accurate predictions, despite the challenge of estimating the noisier third derivative.
Future research will quantify the prediction accuracy of AISE/FS as a function of the effect of the sensor noise on the accuracy of the third derivative of position, that is, the jerk, which is needed to estimate torsion.


\section*{ACKNOWLEDGMENTS}
This research supported by NSF grant CMMI 2031333.

\bibliography{bib_paper,bib_target_tracking}

\begin{thebibliography}{10}

\bibitem{zarchan_book_2012}
P.~Zarchan, {\em Tactical and Strategic Missile Guidance (6th Edition)}.
\newblock American Institute of Aeronautics and Astronautics (AIAA), 2012.

\bibitem{BarShalom2001Estimation}
Y.~Bar-Shalom, X.~Li, and T.~Kirubarajan, {\em Estimation with Applications to Tracking and Navigation: Theory, Algorithms and Software}.
\newblock Wiley, 2001.

\bibitem{zeng_aerospace_aircraft_4D_2022_review}
W.~Zeng, X.~Chu, Z.~Xu, Y.~Liu, and Z.~Quan, ``Aircraft 4d trajectory prediction in civil aviation: A review,'' {\em Aerospace}, vol.~9, no.~2, 2022.

\bibitem{Huang_2022_traj_pred_automobile}
Y.~Huang, J.~Du, Z.~Yang, Z.~Zhou, L.~Zhang, and H.~Chen, ``A survey on trajectory-prediction methods for autonomous driving,'' {\em IEEE Transactions on Intelligent Vehicles}, vol.~7, no.~3, pp.~652--674, 2022.

\bibitem{chatterhi_short_term_1999}
G.~Chatterji, ``Short-term trajectory prediction methods,'' in {\em {AIAA} Guidance, Navigation, and Control Conference}, 1999.

\bibitem{Lymperopoulos_MC_trajPredic_2010}
I.~Lymperopoulos and J.~Lygeros, ``Sequential monte carlo methods for multi-aircraft trajectory prediction in air traffic management,'' {\em International Journal of Adaptive Control and Signal Processing}, vol.~24, no.~10, pp.~830--849, 2010.

\bibitem{ayhan_predictive_analy_2016}
S.~Ayhan and H.~Samet, ``Aircraft trajectory prediction made easy with predictive analytics,'' p.~21–30, Association for Computing Machinery, 2016.

\bibitem{lin_HMM_2018}
Z.~J. . L.~H. Lin~Y., ``An algorithm for trajectory prediction of flight plan based on relative motion between positions,'' in {\em Frontiers Inf Technol Electronic Eng}, pp.~905--916, IEEE Computer Society Press, 2018.

\bibitem{ammoun_rt_prediction_2009}
S.~Ammoun and F.~Nashashibi, ``Real time trajectory prediction for collision risk estimation between vehicles,'' in {\em 2009 IEEE 5th International Conference on Intelligent Computer Communication and Processing}, pp.~417--422, 2009.

\bibitem{Lefkopoulos_MMFK_2021}
V.~Lefkopoulos, M.~Menner, A.~Domahidi, and M.~N. Zeilinger, ``Interaction-aware motion prediction for autonomous driving: A multiple model kalman filtering scheme,'' {\em IEEE Robotics and Automation Letters}, vol.~6, no.~1, pp.~80--87, 2021.

\bibitem{hungu_lee_generalized_1999}
{Lee, H.} and {Tahk, M.}, ``Generalized input-estimation technique for tracking maneuvering targets,'' {\em {IEEE} Transactions on Aerospace and Electronic Systems}, vol.~35, no.~4, pp.~1388--1402, 1999.

\bibitem{shalom-1989-input-est}
Y.~Bar-Shalom, K.~Chang, and H.~Blom, ``{Tracking a Maneuvering Target Using Input Estimation Versus the Interacting Multiple Model Algorithm},'' {\em IEEE Transactions on Aerospace and Electronic Systems}, vol.~25, no.~2, pp.~296--300, 1989.

\bibitem{khaloozadeh_modified_2009}
H.~Khaloozadeh and A.~Karsaz, ``Modified input estimation technique for tracking manoeuvring targets,'' {\em {IET} Radar, Sonar \& Navigation}, vol.~3, no.~1, p.~30, 2009.

\bibitem{gupta_retrospective-cost-based_2012}
R.~Gupta, A.~D'Amato, A.~Ali, and D.~Bernstein, ``Retrospective-cost-based adaptive state estimation and input reconstruction for a maneuvering aircraft with unknown acceleration,'' AIAA 2012-4600, AIAA Guidance, Navigation, and Control Conference, 2012.

\bibitem{ahmed_input_2019}
A.~Ansari and D.~S. Bernstein, ``Input estimation for nonminimum-phase systems with application to acceleration estimation for a maneuvering vehicle,'' {\em IEEE Transactions on Control Systems Technology}, vol.~27, no.~4, pp.~1596--1607, 2019.

\bibitem{han_rcie}
L.~Han, Z.~Ren, and D.~S. Bernstein, ``{Maneuvering Target Tracking Using Retrospective-Cost Input Estimation},'' {\em IEEE Transactions on Aerospace and Electronic Systems}, vol.~52, no.~5, pp.~2495--2503, 2016.

\bibitem{tenne-2002-alpha-beta-gamma}
D.~Tenne and T.~Singh, ``{Characterizing Performance of $\alpha$-$\beta$-$\gamma$ Filters},'' {\em IEEE Transactions on Aerospace and Electronic Systems}, vol.~38, no.~3, pp.~1072--1087, 2002.

\bibitem{hasan-2013-adaptive-alpha-beta}
A.~H. Hasan and A.~Grachev, ``{Adaptive $\alpha$-$\beta$-filter for Target Tracking Using Real Time Genetic Algorithm},'' {\em Journal of Electrical and Control Engineering}, vol.~3, pp.~32--38, 08 2013.

\bibitem{akcal_predictive_2021}
M.~U. Akcal and G.~Chowdhary, ``A predictive guidance scheme for pursuit-evasion engagements,'' in {\em {AIAA} Scitech Forum}, 2021.
\newblock AIAA 2021-1226.

\bibitem{yutian_2020_probalistic}
Y.~Pang and Y.~Liu, {\em Probabilistic Aircraft Trajectory Prediction Considering Weather Uncertainties Using Dropout As Bayesian Approximate Variational Inference}.

\bibitem{bonnabel_cdc_2017_target_tracking}
M.~Pilté, S.~Bonnabel, and F.~Barbaresco, ``Tracking the frenet-serret frame associated to a highly maneuvering target in 3d,'' in {\em 2017 IEEE 56th Annual Conference on Decision and Control (CDC)}, pp.~1969--1974, 2017.

\bibitem{gibbs_fs_iekf_2022}
J.~Gibbs, D.~Anderson, M.~MacDonald, and J.~Russell, ``An extension to the frenet-serret and bishop invariant extended kalman filters for tracking accelerating targets,'' in {\em 2022 Sensor Signal Processing for Defence Conference (SSPD)}, pp.~1--5, 2022.

\bibitem{giulio_2004_frenet}
G.~Avanzini, ``Frenet-based algorithm for trajectory prediction,'' {\em Journal of Guidance Control and Dynamics}, vol.~27, pp.~127--135, 2004.

\bibitem{verma_shashank_2023_realtime_IJC}
S.~Verma, S.~Sanjeevini, E.~D. Sumer, and D.~S. Bernstein, ``{Real-time Numerical Differentiation of Sampled Data Using Adaptive Input and State Estimation},'' {\em International Journal of Control}, pp.~1--13, 2024.

\bibitem{verma_shashank_2024_realtime_VRF_axiv}
S.~Verma, B.~Lai, and D.~S. Bernstein, ``{Adaptive Real-Time Numerical Differentiation with Variable-Rate Forgetting and Exponential Resetting},'' in {\em Proc. Amer. Contr. Conf.}, pp.~3103--3108, 2024.

\bibitem{verma_shashank_ACC2022}
S.~Verma, S.~Sanjeevini, E.~D. Sumer, A.~Girard, and D.~S. Bernstein, ``{On the Accuracy of Numerical Differentiation Using High-Gain Observers and Adaptive Input Estimation},'' in {\em Proc. Amer. Contr. Conf.}, pp.~4068--4073, 2022.

\bibitem{islam2019recursive}
S.~A.~U. Islam and D.~S. Bernstein, ``{Recursive Least Squares for Real-Time Implementation},'' {\em IEEE Contr. Syst. Mag.}, vol.~39, no.~3, pp.~82--85, 2019.

\bibitem{lai2022exponential}
B.~Lai and D.~S. Bernstein, ``{Exponential Resetting and Cyclic Resetting Recursive Least Squares},'' {\em IEEE Contr. Sys. Lett.}, vol.~7, pp.~985--990, 2022.

\bibitem{aastrom1977theory}
K.~J. {\AA}str{\"o}m, U.~Borisson, {\em et~al.}, ``{Theory and Applications of Self-Tuning Regulators},'' {\em Automatica}, vol.~13, no.~5, pp.~457--476, 1977.

\bibitem{malik1991some}
O.~Malik, G.~Hope, and S.~Cheng, ``{Some Issues on the Practical Use of Recursive Least Squares Identification in Self-Tuning Control},'' {\em Int. J. Contr.}, vol.~53, no.~5, pp.~1021--1033, 1991.

\bibitem{mohseni2022recursive}
N.~Mohseni and D.~S. Bernstein, ``{Recursive Least Squares with Variable-Rate Forgetting Based on the {F}-Test},'' in {\em Proc. Amer. Contr. Conf.}, pp.~3937--3942, 2022.

\bibitem{Hanson_FS_formula_1994}
A.~J. Hanson and H.~Ma, ``Visualizing flow with quaternion frames,'' in {\em Proceedings of the Conference on Visualization '94}, p.~108–115, IEEE Computer Society Press, 1994.

\bibitem{hartley_contact_aided_2020}
R.~Hartley, M.~Ghaffari, R.~M. Eustice, and J.~W. Grizzle, ``Contact-aided invariant extended kalman filtering for robot state estimation,'' {\em The International Journal of Robotics Research}, vol.~39, no.~4, pp.~402--430, 2020.

\bibitem{Barfoot_2017}
T.~D. Barfoot, {\em State Estimation for Robotics}.
\newblock Cambridge, 2017.

\bibitem{Chirikjian_2017}
G.~S. Chirikjian, {\em Stochastic Models, Information Theory, and Lie Groups, Volume 2: Analytic Methods and Modern Applications}.
\newblock Springer Science \& Business Media, 2011.

\bibitem{Kalata1983TheTI}
P.~R. Kalata, ``{The Tracking Index: A Generalized Parameter for $\alpha$-$\beta$ and $\alpha$-$\beta$-$\gamma$ Target Trackers},'' {\em The 22nd IEEE Conference on Decision and Control}, pp.~559--561, 1983.

\end{thebibliography}
\bibliographystyle{ieeetr}

\end{document}